\documentclass[a4paper,11pt]{article}
\usepackage{amssymb,amsmath,amsthm,amsfonts}
\usepackage{bookmark}
\usepackage{mathrsfs}
\usepackage{multirow}
\usepackage{graphicx}
\usepackage{authblk}
\usepackage{indentfirst}
\usepackage{multicol}
\usepackage{tabu}
\usepackage{tabularx}
\usepackage{url}
\usepackage{fancyhdr}
\usepackage[square, sort&compress, numbers]{natbib}
\usepackage[all]{xy}
\usepackage{mathtools}
\usepackage[english]{babel}
\usepackage{colortbl}
\usepackage{caption}
\usepackage{setspace}
\usepackage[mathlines]{lineno}
\usepackage{abstract} 
\usepackage{titling}  
\usepackage{mathptmx}  
\usepackage{titlesec}

\titleformat{\section}
{\bfseries\fontsize{12}{16}\selectfont}  
{\thesection}{1em}{}

\titleformat{\subsection}
{\bfseries\fontsize{11}{14}\selectfont}  
{\thesubsection}{1em}{}

\titleformat{\subsubsection}
{\bfseries\fontsize{10}{13}\selectfont}  
{\thesubsubsection}{1em}{}

\titlespacing*{\section}
{0pt}{0pt}{1pt}  
\titlespacing*{\subsection}
{0pt}{0pt}{2pt}
\titlespacing*{\subsubsection}
{0pt}{0pt}{1pt} 

\usepackage[table]{xcolor}
\usepackage{float} 
\usepackage{graphicx} 
\usepackage{booktabs} 
\usepackage{arydshln}
\usepackage{caption} 
\usepackage{array} 
\usepackage[a4paper, margin=1in]{geometry} 
\usepackage{xcolor}
\usepackage{manyfoot}
\definecolor{darkgreen}{rgb}{0.0, 0.5, 0.0} 
\usepackage{float} 
\usepackage{hyperref}
\hypersetup{
	colorlinks=true,
	linkcolor=blue,   
	citecolor=blue,   
	filecolor=blue,   
	urlcolor=blue    
}


\usepackage{newtxtext} 
\usepackage{newtxmath}

\usepackage{amssymb}    
\usepackage{amsmath}   
\usepackage{microtype} 
\usepackage{geometry} 
\usepackage{hyperref}
\usepackage{xr-hyper}
\usepackage{cleveref}   
\usepackage[labelfont=bf]{caption}

\begin{document}
\title{Persistent Directed Flag Laplacian (PDFL)-Based Machine Learning for Protein–Ligand Binding Affinity Prediction}

\vspace{1em}

\author{
	Mushal Zia$^{1}$, 
	Benjamin Jones$^{1}$, 
	Hongsong Feng$^{1}$, 
	and Guo-Wei Wei$^{1,2,3, \dagger}$  \\
$^1$  Department of Mathematics, \\
 Michigan State University, East Lansing, MI 48824, USA. \\
$^2$ Department of Electrical and Computer Engineering, \\
 Michigan State University, East Lansing, MI 48824, USA. \\
 $^3$ Department of Biochemistry and Molecular Biology, \\
 Michigan State University, East Lansing, MI 48824, USA. \\
 $^\dagger$ Address correspondences to Guo-Wei Wei. E-mail: weig@msu.edu}

\date{}

\maketitle  

\newpage

\vspace{1em}  

\begin{abstract}   
Directionality in molecular and biomolecular networks plays a significant role in the accurate represention of the complex, dynamic, and asymmetrical nature of interactions present in protein-ligand binding, signal transduction, and biological pathways. Most traditional techniques of topological data analysis (TDA), such as persistent homology (PH) and persistent Laplacian (PL), overlook this aspect in their standard form. To address this, we present the persistent directed flag Laplacian (PDFL), which incorporates directed flag complexes to account for edges with directionality originated from polarization, gene regulation, heterogeneous interactions, etc. This study marks the first application of the PDFL, providing an in-depth analysis of spectral graph theory combined with machine learning. Besides its superior accuracy and reliability, the PDFL model offers simplicity by requiring only raw inputs without complex data processing. We validated our multi-kernel PDFL model for its scoring power against other state-of-art methods on three popular benchmarks, namely PDBbind v2007, v2013, and v2016. Computational results indicate that the proposed PDFL model outperforms competitors in protein-ligand binding affinity predictions, indicating that PDFL is a promising tool for protein engineering, drug discovery, and general applications in science and engineering.     
\end{abstract}

\noindent \textbf{Keywords}    
  Persistent topological Laplacians,  Directed flag Laplacians, Clique complex, Protein-ligand binding, Machine learning.
\vspace{2.5em}   


\newpage
\section{Introduction}

Protein-ligand binding is a crucial interaction in which a protein, such as an enzyme or a receptor, binds to one or more ligands. This interaction is fundamental to various biological processes, including cell signaling, molecular transport, and metabolism. While proteins typically bind to only certain ligands due to their specific structural, electrostatic, and chemical compatibility as in hormone-receptor binding and enzyme-substrate interactions, there are cases a protein binds to multiple small molecules. Protein-ligand binding is typically governed by non-covalent forces, including hydrogen bonds, van der Waals forces, hydrophobic interactions, electrostatic interactions, and ionic bonds. Under the physiological condition, hydrophobic residues on the protein interact with non-polar ligand to stabilize the binding. In drug discovery, many pharmaceuticals are designed to bind to specific proteins, like receptors or enzymes, to modulate their functions. Moreover, protein-ligand interactions are also exploited in biosensors, diagnostic strategies, and protein engineering.

Experimental studies on protein-ligand binding are expensive. Computer modeling of protein-ligand binding is essential for understanding molecular interactions and plays a significant role in drug discovery. Various computational methods have been developed for predicting the binding modes, affinities, and thermodynamic properties of these interactions. Among them, machine learning predictions have become very popular in the past decade \cite{ballester2010machine, pan2022aa, wang2017improving} due to their ability to incorporate fast-growing experimental data for accurate predictions. Mathematical artificial intelligence (Math AI) paradigms \cite{nguyen2019mathematical, nguyen2020mathdl}, such as  topological deep learning (TDL) \cite{cang2017topologynet}, has been introduced for protein-ligand binding prediction \cite{cang2018representability, cang2018integration}. This approach secured some of the best results in D3R Grand Challenges, an annual worldwide competition series in computer-aided drug design \cite{nguyen2019mathematical, nguyen2020mathdl}.

One of the key components of TDL is persistent homology (PH) \cite{edelsbrunner2000topological, zomorodian2004computing}, an algebraic topology tool for topological data analysis (TDA). It is well recognized that the foundational idea behind TDA is that data exhibit a discernible shape regardless of its complexity, and that shape matters. In other words, the topological or geometric structure of the data can reveal significant key patterns and relationships that may not be instantly recognizable from raw data alone. TDA utilizes PH to study and track features across varying scales by merging concepts from algebraic topology and data science, thus revealing insights into the connectivity and geometry of the underlying data that cannot be obtained from traditional mathematical, statistical, and physical methods \cite{cheng2009comparative}. The strength of TDA lies in its capability to encompass essential features through topological signatures while filtering out irrelevant ones, thus allowing the simplification of data into a more fitting form for analysis. These signatures are succinct mathematical representation of topological features such as connectivity, loops, voids, and higher-dimensional analogues, which capture the shape and underlying organization of complex data. Betti numbers count the number of these topological features in various dimensions; for instance, a loop captured by a one-dimensional Betti number indicates the presence of a closed pathway or feedback cycle in space. Similarly, a persistent Betti number provides information about the evolution of these topological features across varying scales.

While PH and other algebraic topological techniques have successfully helped simplify complex molecular structures and reduce dimensions in biological systems \cite{nguyen2020review}, these tools struggle with tracking non-topological shape evolution. The persistent combinatorial Laplacian, or persistent Laplacian (PL), was introduced in 2019 to address some of PH's limitations \cite{wang2020persistent}. PL can be regarded as a generalization of the classical graph Laplacian to higher topological dimensional complex structures and as an extension of PH to the non-harmonic spectra, as its harmonic spectra recover PH's topological invariants. Moreover, PL may be considered as the counterpart of the persistent Hodge Laplacian, which is defined on differentiable manifolds \cite{chen2019evolutionary}. The persistent Hodge Laplacian uses differential forms, rather than simplicial complexes, to study and analyze manifolds, cohomology, and topological invariants in higher-dimensional spaces. PL has stimulated much theoretical interest \cite{ memoli2022persistent,liu2023algebraic,wang2023persistent,dong2024faster} and had significant application \cite{meng2021persistent}, including protein engineering \cite{qiu2023persistent} and the accurate forecasting of emerging dominant virus variants \cite{chen2022persistent}. A review of this trending topic is available in literature \cite{wei2023persistent}.

In a further advancement, following the framework of persistent directed flag complex homology \cite{lutgehetmann2020computing}, Jones and Wei have  recently proposed a new TDA tool called the persistent directed flag Laplacian (PDFL) \cite{jones2023persistent}. Directed flag Laplacian neglects multiscale analysis and thus, has limited utility for data analysis.  PDFL extends the concept of persistent Laplacian to a persistent directed flag complex, offering a new topological tool for modeling gene regulation, directed graph, atomic polarization, etc. This extension allows for a more comprehensive analysis, particularly in directed systems, by introducing a spatial dimension through the filtration parameter. The directed flag Laplacian provides a deeper analysis of multiscale topological features in directed networks by acting on simplicial complexes and directed graphs (digraphs). Moreover, it expands TDA's capability by incorporating comprehensive insights into complex molecular structures, thus, enabling the handling of directional network data, which can be challenging at times.

In this work, we present the first application of PDFL by integrating spectral graph theory with flexibility-rigidity index (FRI)-based methods \cite{xia2013multiscale}, enabling advanced machine learning techniques such as gradient boost decision trees (GBDT) with topological atomic descriptors. This framework accounts for the directionality of protein-ligand interactions based on their electronegativity differences and is used to develop multi-kernel predictive models tested against three large datasets. In the following sections, we explore the mathematical foundations, computational implementation, and the impressive performance of our PDFL model across multiple benchmark datasets.

\begin{figure}[htbp]
	\centering
	\includegraphics[width=\textwidth]{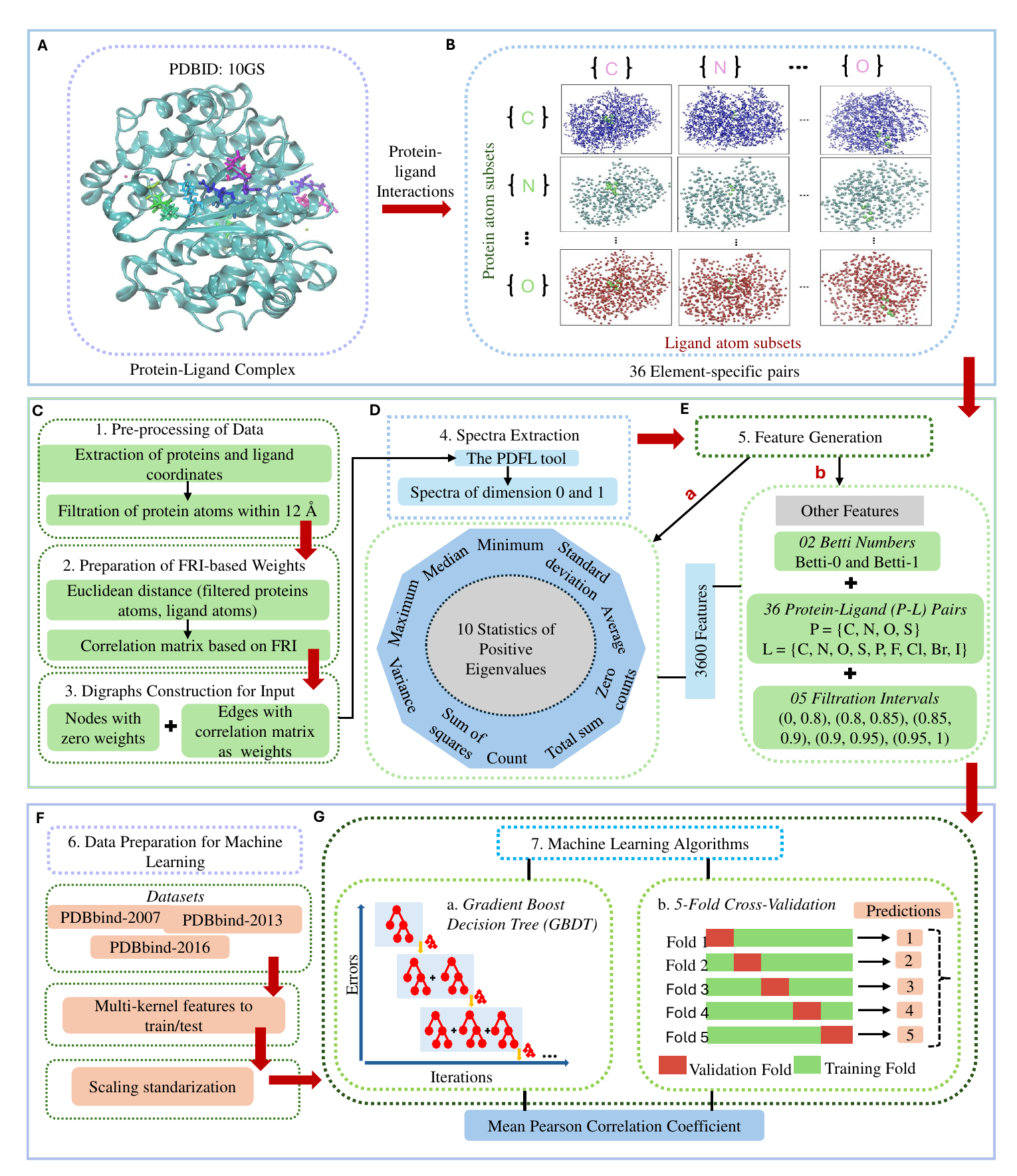}
	\caption{Illustration of workflow for protein-ligand binding affinity prediciton through Persistent Directed Flag Laplacian (PDFL) learning. (A) A protein-ligand complex with PDBID 10GS is shown. (B) A graphical potrayal of 36 element-specific pairs formed by four protein atoms and 9 ligand atoms representing their interactions. (C-E) An outline of data processing and input preparation for PDFL to generate spectra is shown. A total of 3600 features are generated for each protein-ligand complex by a combination of 36*[Element-types]*5[Intervals]*2[Betti numbers]*10[Statistical features]. (F, G) Three datasets from the Protein Data Bank (PDB) are used in this work to train the PDFL model with machine learning techniques, such as Gradient Boost Decision Tree (GBDT) and 5-fold Cross-Validation. Mean Pearson correlation coefficient is calculated  for a number of predictions which shows the strength of protein-ligand binding affinity.}
	\label{fig:flowchart}   
\end{figure}

 \vspace{1.5em}   
\section{Results and Discussion} 
  \vspace{0.5em}     
\subsection{ An Overview to the Persistent Directed Flag Laplacian Learning  }

The persistent directed flag Laplacian (PDFL) is built by extending  the software Flagser \cite{lutgehetmann2020computing} and leveraging its core functionalities to generate directed flag complexes and the corresponding (co)boundary matrices. By taking weighted and filtered digraphs as an input, PDFL computes the spectra across a range of filtration values to capture multiscale topological features of the digraphs. The input digraphs encompass both the directionality and strength of atomic interactions, with edge directions determined by electronegativity differences and weights derived from the correlation matrix (see \textcolor{blue}{\autoref{fig:flowchart}C}). These properties make our digraphs well-suited for spectral analysis, providing information about the connectivity and flow within the protein-ligand interaction network. The filtration parameter in PDFL varies independently from these edge weights. We specify a filtration range from 0 to 1, incremented by 0.01, and at each filtration level \( \epsilon \), PDFL computes the spectra of Laplacian matrices for the subgraph corresponding to edges with filtration values below \( \epsilon \). This multiscale spectral analysis across multiple filtration levels allows us to identify significant patterns in the connectivity and structure of the networks, which may not be visible when considering the entire graph at once. This tool computes the spectra of Laplacian matrices associated with directed flag complexes constructed from our filtered digraphs (see \textcolor{blue}{\autoref{fig:flowchart}D}). A detailed analysis of model evaluation metrics, selection of data from Protein Data Bank (PDB) for featurization, and machine learning algorithms used for training our PDFL models are discussed in Section \textcolor{blue}{\hyperref[sec:SI]{SI}}.
 
 \vspace{0.5em}      
 \subsubsection{Statistical Features Extraction}
 
 To capture the topological characteristics of our protein-ligand interaction networks, we compute a set of descriptive statistics from the spectral eigenvalues, including the minimum non-zero eigenvalue (the Fiedler value for Laplacian matrices), the maximum eigenvalue, and the sum, mean, median, variance, and standard deviation of all positive eigenvalues. Furthermore, we also incorporate the count of positive eigenvalues, the sum of squares, and the zero count, which represents the number of eigenvalues nearly equal to zero. The analysis is performed over the set of filtration intervals:  \((0, 0.8)\), \((0.8, 0.85)\), \((0.85, 0.9)\), \((0.9, 0.95)\), and \((0.95, 1)\), chosen to provide a comprehensive view of the evolving complexity of simplices. We find that the key topological features, Betti-0 and Betti-1, appear to be more concentrated in above defined intervals. Moreover, we consider four protein atoms, \{C, N, O, S\} and nine ligand atoms, \{C, N, O, S, P, F, Cl, Br, I\}. Hydrogen (\text{H}) atom is excluded to reduce complexity. This results in a total of 3600 = 36*[Element-types]*5[Intervals]*2[Betti numbers]*10[Statistical features]  for each protein-ligand complex within a cutoff value of 12 \AA. A detailed workflow is demonstrated in \textcolor{blue}{\autoref{fig:flowchart}E}.

  \vspace{0.5em}      
 \subsubsection{Hyperparameters  }
 
For the quantitative prediction of protein-ligand binding affinities, we built our model by integrating the rigidity index from equation \eqref{eq:10} with GBDT. As listed in \autoref{tab:param}, various scales are chosen for the values of \(\beta\) (kernel order) and \(\tau\) (decay parameter) for the efficient representation of protein-ligand binding interaction strength. Moreover, we refer to the model as PDFL\(^\Omega_{\beta,\tau}\) with \(\beta = \nu\) when \(\Omega = L\), the generalized Lorentz kernel, and \(\beta = \kappa\) when \(\Omega = E\), the generalized exponential kernel. 
 
  \vspace{1em}     
 \newlength{\tablewidth}
 \setlength{\tablewidth}{\dimexpr\linewidth-2cm\relax} 
 \setlength{\textfloatsep}{15pt}  
 
 \begin{table}[htbp!]
 	\centering
 	\hspace*{0cm} 
 	\textcolor{darkgreen}{\hspace*{-0.09cm}\rule{\dimexpr\tablewidth+0.4cm\relax}{4pt}} 
 	
 	\vspace{-1pt} 
 	
 	\noindent\colorbox{gray!15}{ 
 		\begin{minipage}{\tablewidth} 
 			\captionsetup{justification=centering, font=small, labelfont=bf, skip=5pt} 
 			\caption{\textbf{\fontsize{9}{11}\selectfont Various Scales of Hyperparameters for PDFL Model.} Here, \(\beta = \nu\) for \(\Omega = L\), the generalized Lorentz kernel, and \(\beta = \kappa\) when \(\Omega = E\), the generalized exponential kernel.} 
 			\label{tab:param} 
 			\vspace{2pt} 
 			
 			\centering
 			\small
 			\renewcommand{\arraystretch}{1.5} 
 			\setlength{\tabcolsep}{10pt} 
 			\begin{tabular}{c c} 
 				\rowcolor{gray!15} 
 				\textbf{\fontsize{9}{11}\selectfont Parameters} & \textbf{\fontsize{9}{11}\selectfont Domain of Values} \\
 				\midrule
 				\(\beta\) & \{0.5, 1.0, 1.5,..., 6\} \\
 				\hdashline[1pt/1pt] 
 				\(\tau\) &  \{0.5, 1.0, 1.5,..., 6\} $\cup$ \{10, 15, 20\} \\
 			\end{tabular}
 			
 			\vspace{4pt} 
 		\end{minipage}
 	}
 	
 	\vspace{0pt}
 \end{table}
 
 \vspace{1em}     
 \subsubsection{Data Preparation for Machine Learning}
 
 We validate our PDFL-based machine learning model using three benchmark datasets from the Protein Data Bank (PDB) database \cite{liu2015pdb,berman2000protein}: PDBbind v2007  \cite{cheng2009comparative}, PDBbind v2013 \cite{li2014comparative}, and PDBbind v2016  \cite{liu2017forging} which are widely used to assess the general performance of scoring functions on diverse protein-ligand complexes \cite{cheng2009comparative, li2014comparative, liu2017forging, nguyen2019agl, cang2017analysis, ballester2010machine, wang2017improving, li2013id, li2015improving, nguyen2019dg, meng2021persistent, wee2021forman,liu2021persistent}. Details of the three datasets are given in \autoref{tab:datasets}. Furthermore, \autoref{tab:parameters} provides the detailed setting of parameters for our machine learning model (GBDT). For further validation, we also employ 5-fold cross-validation to optimize the kernel hyperparameters \(\Omega\), \(\beta\), and \(\tau\) using the same parameters for consistency and robustness. We notice that minor variations in these hyperparameters have a little to no effect on the overall prediction accuracy.

 \vspace{1em}    
\subsection{Standard PDFL Models for Binding Affinity Predictions}

We begin by validating the scoring power of our model by employing the PDBbind v2007 core set, consisting of 195 protein-ligand complexes across 65 clusters as our test set. The training data (N = 1105) is compiled based on the refined set which contains 1300 complexes excluding the core set (see  \autoref{tab:datasets}). Moreover, we denote \(R_p^m\) and \(R_p^b\) as the mean and best Pearson correlation coefficients, respectively, and similarly, the mean and best cross-validation scores are represented by \(CV^m\) and \(CV^b\). For the generalized Lorentz kernel model PDFL\(^L_{3,2}\), the Pearson correlation coefficient \((R_p^m, R_p^b) = (0.828, 0.831)\) is achieved with an RMSE of 1.370 kcal/mol, alongside a 5-fold cross-validation score of \((CV^m, CV^b) = (0.743, 0.783)\). Another model, PDFL\(^L_{3,1}\), performs similarly well, with \((R_p^m, R_p^b) = (0.825, 0.829)\) and an RMSE of 1.395 kcal/mol, while the cross-validation scores are \((CV^m, CV^b) = (0.743, 0.768)\). Furthermore, PDFL\(^L_{3.5,1.5}\) also produces strong results, with a Pearson correlation coefficient of \((R_p^m, R_p^b) = (0.824, 0.827)\) and an RMSE of 1.381 kcal/mol, together with cross-validation scores of \((CV^m, CV^b) = (0.745, 0.768)\). Among the generalized exponential kernel, PDFL\(^E_{10,3.5}\) manages to reach comparable performance, with \((R_p^m, R_p^b) = (0.823, 0.827)\) and an RMSE of 1.397 kcal/mol, with cross-validation scores of \((CV^m, CV^b) = (0.731, 0.774)\). Similarly, the another model PDFL\(^E_{6,2.5}\) shows \((R_p^m, R_p^b) = (0.818, 0.821)\) and an RMSE of 1.411 kcal/mol, with cross-validation scores of \((CV^m, CV^b) = (0.738, 0.772)\). Both the models PDFL\(^L_{15,3}\) and PDFL\(^L_{1.5,0.5}\) stand very close to PDFL\(^E_{10,3.5}\), with mean and best Pearson correlation coefficients of 0.823 and 0.827 for both models. PDFL\(^L_{15,3}\) recorded cross-validation scores of 0.736 and 0.773, with an RMSE of 1.386 kcal/mol, while PDFL\(^L_{1.5,0.5}\) has slightly higher RMSE at 1.402 kcal/mol, and its cross-validation scores are 0.737 and 0.765.  \autoref{fig:heatmap1} illustrated the impact of mean Pearson correlation coefficients ($R_p$) of best performing one-scale PDFL\(^\Omega_{\beta,\tau}\) models for the generalized Lorentz kernel and generalized exponential kernel. Among all, PDFL\(^L_{3,2}\), PDFL\(^L_{3,1}\), and PDFL\(^E_{10,3.5}\) outperform the rest and emerge as the best-performing models with consistently higher correlation coefficient and cross-validation scores (see \textcolor{blue}{\autoref{fig:heatmap2}A}). In contrast, the Lorentz kernel model PDFL\(^L_{15,0.5}\) with \(v=15\) and \(\tau=0.5\) performs the worst, with \((R_p^m, R_p^b) = (0.7807, 0.783)\), an RMSE of 1.511 kcal/mol, and relatively lower cross-validation scores of \((CV^m, CV^b) = (0.694, 0.745)\). A full comparison of all tested models and results is listed in \autoref{tab:Kernel_Results} and their impact is demonstrated in \textcolor{blue}{\autoref{fig:heatmap4}A}. Moreover, \autoref{fig:heatmap3} represents the mean $R_p$ of the 5-fold cross-validation experiments for the best performing one-scale generalized Lorentz kernel and generalized exponential kernel models, while a combined impact of both kernels can be seen in \textcolor{blue}{\autoref{fig:heatmap2}B}.

\begin{figure}[t!]
	\centering
	\includegraphics[width=\textwidth]{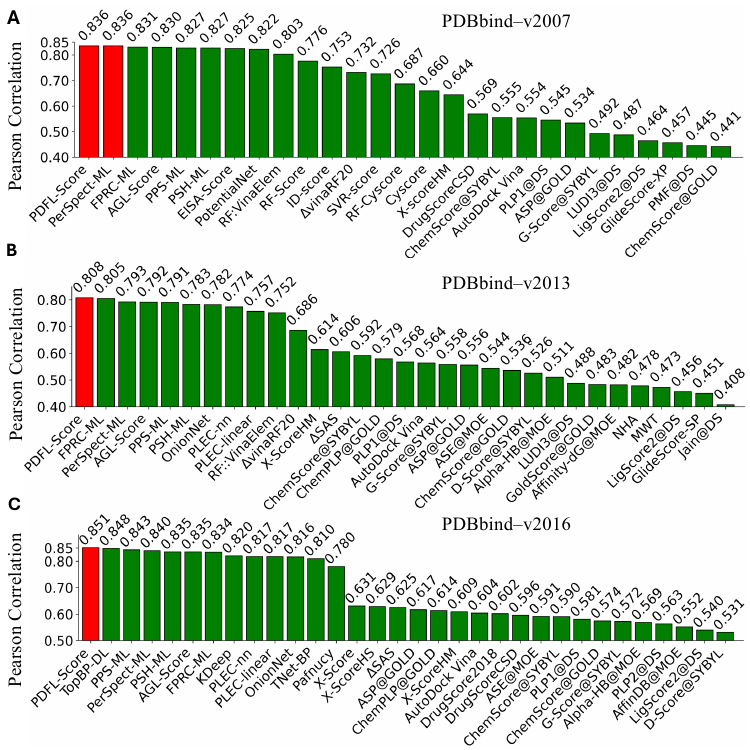}
	\vspace{-15pt}
	\caption{ Performance comparison of our PDFL model for three well-established protein-ligand binding affinity datasets, including  (A)  PDBbind-2007: Our model attains same value of pearson correlation coefficient $R_p$ = 0.836 as the PerSpect-ML model \cite{meng2021persistent} (both highlighted in red). (B)  PDBbind-2013: PDFL-Score outperforms all other methods (highlighted in green), including the FPRC-ML Score with $R_p$ = 0.805 \cite{wee2021forman} (C) PDBbind-2016: PDFL achieved an $R_p$ = 0.851 (highlighted in red). These results across three datasets validates the robustness and accuracy of the predictive power of PDFL.}
	\vspace{-4pt} 	    
	\label{fig:barchart}
	 \vspace{10pt}   
\end{figure}

We further extend our work to incorporate two-kernel models as they have shown to improve the accuracy of binding free energy predictions in the previous studies \cite{cang2018integration,xia2013multiscale,nguyen2017rigidity,nguyen2019agl, meng2021persistent} and generate a total of 7200 features against each protein-ligand complex. Among the two-kernel PDFL\(^{\Omega_1\Omega_2}_{\beta_1,\tau_1; \beta_2,\tau_2}\) models tested, the Lorentz-Lorentz configuration PDFL\(^{LL}_{3,2; 3,1}\) delivers the best performance, with \((R_p^m, R_p^b) = (0.830, 0.836)\) and an RMSE of 1.368 kcal/mol. Close behind, the mixed Lorentz-exponential model PDFL\(^{LE}_{3,2; 10,3.5}\) achieves \((R_p^m, R_p^b) = (0.827, 0.832)\) and an RMSE of 1.383 kcal/mol. Another mixed kernel model, PDFL\(^{LE}_{3,1; 10,3.5}\), shows strong results with \((R_p^m, R_p^b) = (0.829, 0.834)\), accompanied by an RMSE of 1.381 kcal/mol. Furthermore, the Lorentz-exponential model PDFL\(^{LE}_{3,2; 6,2.5}\) and PDFL\(^{LE}_{3,1; 6,2.5}\) exhibits Pearson correlation values of \((R_p^m, R_p^b) = (0.826, 0.830)\) and \((R_p^m, R_p^b) = (0.824, 0.827)\), respectively. The best model among the exponential-exponential configuration is found to be PDFL\(^{EE}_{10,3.5; 6,2.5}\), with an RMSE of 1.398 kcal/mol and \((R_p^m, R_p^b) = (0.824, 0.826)\). It is evident from these results that the two-scale predictions consistently outperform the single-scale models. Moreover, it becomes apparent that models incorporating generalized Lorentz kernels outperform the models that solely comprises of generalized exponential kernels. Finally, the best-performing two-kernel model among the Lorentz-Lorentz configuration is PDFL\(^{LL}_{3,2; 3,1}\), while the highest score among the  exponential-exponential configuration is gained by PDFL\(^{EE}_{10,3.5; 6,2.5}\) (see  \textcolor{blue}{\autoref{fig:heatmap4}B}).

Based on a comparatively better predictive performance of a two-kernel model than a single-kernel model, we now explore the possibility of utilizing four-kernel PDFL models. In this analysis, a total of 14400 features are extracted for each complex, generated by four corresponding kernels by ensuring no overlap in parameter configuration. This approach allows for a more comprehensive evaluation of multiscale interactions across distinct kernel parameter spaces. Furthermore, we focus on specific sets of \(\beta\) and \(\tau\) since the entire parametric evaluation is excessively expensive. For the prediction of the PDBbind v2007 core set, feature vectors having zero values for all complexes are ignored which gives rise to a total of 14373 features. We find PDFL\(^{LLEE}_{3,2; 3,1;10,3.5; 6,2.5}\) to be the best performing model out of all with a mean Pearson correlation of 0.830 and a best value of 0.833 with the RMSE value of 1.370 kcal/mol. 
 \autoref{tab:BA2007}  enlists all PDBID, experimental and predicted binding affinities generated by PDFL\(^{LLEE}_{3,2; 3,1;10,3.5; 6,2.5}\) model on PDBbind v2007 core set.

\begin{figure}[t!]
	\centering
	\vspace{-6mm}  
	\includegraphics[width=0.8\linewidth]{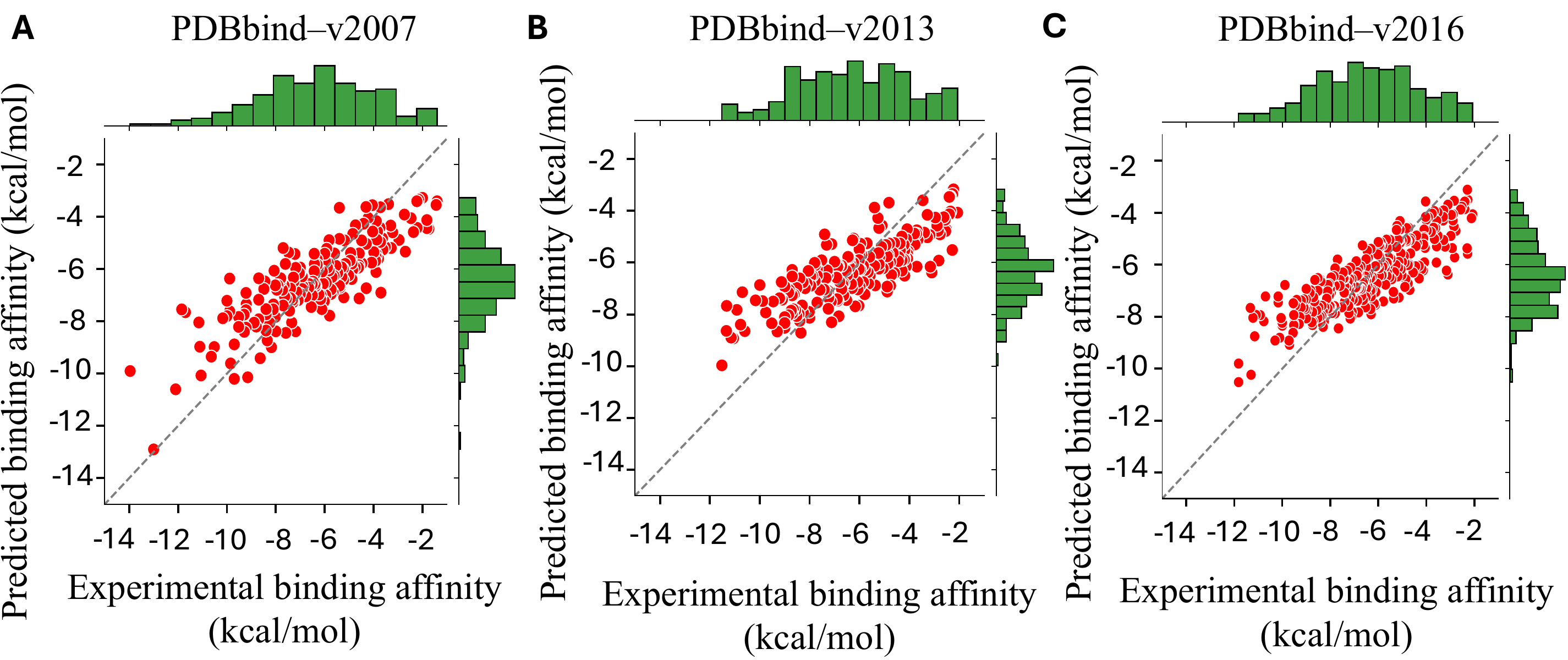}
	\caption{ Scatterplot comparison  of experimental binding affinities and prediction results for  (A)  PDBbind-v2007,  (B)  PDBbind-v2013, and  (C)  PDBbind-v2016. The corresponding Pearson correlation coefficient values are 0.836, 0.808, and 0.851, respectively.}
	\label{fig:scatterplot}
	 \vspace{10pt}   
 
\end{figure}

 \vspace{1em}    
\subsection{Consensus PDFL Models for Binding Affinity Predictions}

Finally, we develop a consensus PDFL model to further improve the predictive performance by 
integrating the features generated from our final PDFL\(^{LLEE}_{3,2; 3,1;10,3.5; 6,2.5}\) model and a pre-trained transformer-based predictive model, known to as model-seq. This type of consensus prediction has been performed in a recent study \cite{feng2024mayer}. Model-seq leverages molecular features using amino acid sequences and SMILES strings data as an input to generate molecular embeddings for proteins \cite{rives2021biological}, and small molecules \cite{winter2019learning}. These embeddings are, then, integrated with GBDT algorithm to enhance the predictive accuracy of complex protein-ligand interaction framework.

\begin{table}[t]
	
	\textcolor{darkgreen}{\noindent\rule{\dimexpr\textwidth+1.7\fboxsep\relax}{4pt}} 
	
	\vspace{-1pt} 
	
	\setlength{\fboxsep}{0pt} 
	\noindent\hspace*{-\fboxsep}\colorbox{gray!15}{ 
		\begin{minipage}{\dimexpr\textwidth+2\fboxsep\relax} 
			
			
			\vspace{8pt} 
			
			\captionsetup{justification=centering, font=small, labelfont=bf, skip=5pt} 
			\caption{  The Pearson correlation coefficients and root mean square errors (RMSE) (in kcal/mol) of the three consensus models.   First column: The three test sets of PDBbind-2007, PDBbind-2013, and PDBbind-2016.  Second column: PDFL-Scores along with their corresponding RMSEs.  Third column: Transformer-model scores along with the corresponding RMSEs values.  Fourth column: The consensus scores of PDFL + Transformer model for the three datasets.} 
			\label{tab:concensus} 
			\vspace{2pt} 
			
			\centering
			\renewcommand{\arraystretch}{1.3} 
			\setlength{\tabcolsep}{0pt} 
			\begin{tabular*}{\textwidth}{@{\extracolsep{\fill}} l c c c} 
				\rowcolor{gray!15} 
				
				\textbf{\fontsize{9}{11}\selectfont } & \textbf{\fontsize{9}{11}\selectfont PDFL} & \textbf{\fontsize{9}{11}\selectfont Transformer} & \textbf{\fontsize{9}{11}\selectfont PDFL + Transformer} \\
				\midrule 
				\fontsize{9}{11}\selectfont PDBbind-2007 & \fontsize{9}{11}\selectfont 0.830 (1.370) & \fontsize{9}{11}\selectfont0.795(2.006) & \fontsize{9}{11}\selectfont 0.836 (1.374) \\
				\hdashline[1pt/1pt] 
				\fontsize{9}{11}\selectfont PDBbind-2013 & \fontsize{9}{11}\selectfont 0.770 (1.494) & \fontsize{9}{11}\selectfont 0.791(1.977) & \fontsize{9}{11}\selectfont 0.808 (1.435) \\
				\hdashline[1pt/1pt] 
				\fontsize{9}{11}\selectfont PDBbind-2016 & \fontsize{9}{11}\selectfont 0.819 (1.326) & \fontsize{9}{11}\selectfont0.836(1.716) & \fontsize{9}{11}\selectfont 0.851 ( 1.252) \\
				
			\end{tabular*}
			
			\vspace{0pt} 
		\end{minipage}
	}
	 \vspace{10pt}   
\end{table}

It remains to be shown that the outstanding performance of the PDFL model is not limited to a specific dataset. We now consider PDBbind v2013 \cite{li2014comparative} and train our model on the refined set (\(N = 2959\)) excluding its core set (\(N = 195\)) which we use as our test set. With total of 14284 statistical features, our PDFL-score comes out to be \((R_p^m, R_p^b) = (0.77, 0.777)\) with an RMSE value of 1.494 kcal/mol. PDBbind v2016 is the last benchmark considered in this study which consists of 290 protein-ligand complexes in 58 clusters \cite{liu2017forging}. We train our PDFL model on its refined set (\(N=4075\)), excluding its core set. Both training and test data of this dataset is larger than its predecessor, PDBbind v2013 with its PDFL-score even better. From a total of 14384 features, the Pearson correlation coefficient \((R_p^m, R_p^b) = (0.819, 0.821\)) is achieved with an RMSE of 1.326kcal/mol. Finally, by integrating the 20 best predictions from all three benchmarks with those from the 20 model-seq predictions into what we refer to as the consensus (PDFL + Transformer) model, we achieve Pearson correlation coefficient values of 0.836, 0.808, and 0.851, respectively. In \autoref{tab:concensus}, mean values of the scoring functions of the PDFL model, Transformer model, and the consensus model (PDFL + Transformer) are provided in the the second, third, and fourth column, respectively, along with their corresponding RMSEs. 

In the first benchmark PDBbind-v2007, our PDFL model attains same value of pearson correlation coefficient $R_p$ = 0.836 as the PerSpect-ML model with an RMSE of 1.907 kcal/mol \cite{meng2021persistent}. The runner-up is FPRC-ML model \cite{wee2021forman} with the an $R_p$ of 0.831. \textcolor{blue}{\autoref{fig:barchart}A} shows the performance comparison of the two models along with a number of scoring functions obtained in the earlier studies. As for the second benchmark PDBbind-v2013, the proposed model is able to outperform all the previous scoring functions, including the FPRC-ML \cite{wee2021forman}, PerSpect-ML \cite{meng2021persistent}, and the AGL \cite{nguyen2019agl} model with an $R_p$ = 0.808 (See \textcolor{blue}{\autoref{fig:barchart}B}). The recorded value of root mean square error (RMSE) is 1.435 kcal/mol. A full comparison of experimental and predicted binding affinities is presented in \autoref{tab:BA2013} along with the corresponding PDBIDs.

For PDBbind-v2016, our PDFL model achieves an outstanding $R_p$ value of 0.851 with an RMSE = 1.252 kcal/mol. We compare the performance of our PDFL model with various models as shown in  \textcolor{blue}{\autoref{fig:barchart}C}. Moreover, a detailed comparison of the experimental and predicted binding affinities generated by PDFL\(^{LLEE}_{3,2; 3,1;10,3.5; 6,2.5}\) is enlisted in \autoref{tab:BA2016}. Furthermore, a correlation comparison of experimental binding affinities and prediction results for the three datasets is illustrated in \textcolor{blue}{\autoref{fig:scatterplot}}. These results validate the reliability and accuracy of the predictive power of PDFL.

 \vspace{1.5em}   
\section{Methods and Algorithms}
  \vspace{0.5em}     
\subsection{Mathematical Foundations of the Persistent Directed Flag Laplacian}

\setlength{\parskip}{0em} 
One of the basic topological representation includes combinatorial structures called simplicial complexes, which are composed of simplices and used as a generalization of graphs or networks. Mathematically, a $k$-simplex $\sigma^{k}=\left\{w_{0}, w_{1}, \ldots, w_{k}\right\}$ is the convex hull of affinely independent points $\left\{w_{i}\right\}_{i=0}^{k} \subset \mathbb{R}^{k}$, defined as follows:

$$\sigma^{k}=\left\{\sum_{i=0}^{k} \lambda_{i} w_{i} \mid \lambda_{i} \geq 0 \text { for all } i ; \sum_{i=0}^{k} \lambda_{i}=1\right\}.$$

\begin{figure}[h!]
	\centering
	\includegraphics[width=0.8\textwidth]{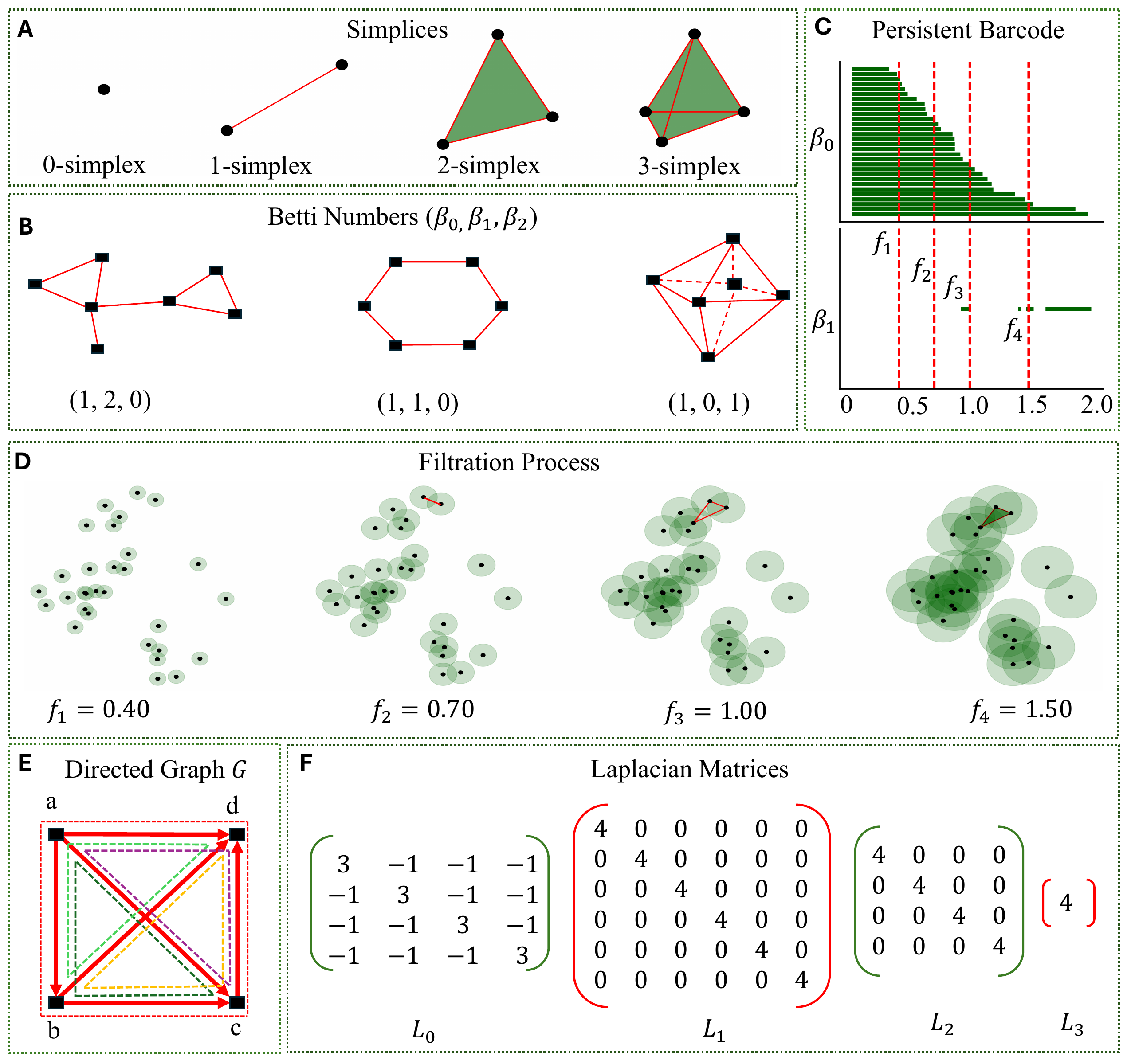}
	\caption{Illustration of fundamental concepts in TDA.  (A) A $k$-simplex is made from $k+1$ vertices and can ve viewed, geometrically, as a point $(0$-simplex) defined by exactly 1 vertex, an edge (1-simplex) determined by 2 connected vertices, a triangle (2-simplex) formed by 3 vertices and 3 edges, and a tetrahedron made up of 3-simplex. (\textbf{B}) Representation of Betti numbers $\beta_{k}$. Geometrically, $\beta_{0}$ is the number of connected components, $\beta_{1}$ is the number of tunnels, loops, or circles, while $\beta_{2}$ represents the number of voids or cavities.  (\textbf{C}) Visualization of a persistence barcode rendered from the filtration process. (\textbf{D}) Demonstration of a filtration process where simplical complexes are formed by interactions at four distinct filtration values denoted by scales $f_1$, $f_2$, $f_3$, and $f_4$. (\textbf{E}) The directed graph (digraph ) \( G\), a minimal example of a directed 4-clique, with 4 vertices \( V = \{a, b, c, d\} \) (all shown in black), 6 edges \( E = \{(a, b), (a, c), (a, d), (b, c), (b, d), (c, d)\} \) (all shown in bold red lines with arrows), 4 \( 3 \)-cliques \( \{(a, b, c) , (a, b, d), (a, c, d), (b, c, d)\} \) (shown in four different dotted colored lines), and the 4-clique \( (a, b, c, d) \) (shown in dotted red lines). (\textbf{F}) Generation of Laplacian matrices from the graph and directed flag complex of \( G \). The spectra of the corresponding matrices \( L_0 \), \( L_1 \), \( L_2 \), and \( L_3 \) are computed as \(\{0, 4, 4, 4\}\), \(\{4, 4, 4, 4, 4, 4\}\), \(\{4, 4, 4, 4\}\), and \(\{4\}\), respectively, along with the Betti numbers \( \beta_0 = 1 \), \( \beta_1 = 0 \), \( \beta_2 = 0 \), and \( \beta_3 = 0 \).}
	\vspace{-7pt} 
	\label{fig:simplexes}  
	\vspace{10pt}   
\end{figure}

Here, the convex hull formed by any subset of $j+1$ vertices from the $k+1$ points $\left\{w_{0}, w_{1}, \ldots, w_{k}\right\}$ is called a $j$-dimensional face of $\sigma^{k}$, where $j$ ranges from 0 to $k-1$. It can be geometrically viewed as a point $(0$-simplex, 0-clique, and a 0-dimensional face) defined by exactly 1 vertex, an edge (1-simplex, 1-clique, and a 1-dimensional face) determined by 2 connected vertices, a triangle (2-simplex, 3-clique, and a 2-dimensional face) formed by 3 vertices and 3 edges, a tetrahedron made up of (3-simplex, 4-clique, and a 3-dimensional face) defined (\textcolor{blue}{\autoref{fig:simplexes}A}).

Homology provides a way to study the topological properties of a space by associating algebraic structures with its geometric features, using tools such as chain groups and boundary operators. $C_{k}$ is the formal span over $k$-simplices, so $C_{0}$ is a vector space with vertices as its basis, $C_{1}$ with edges as its basis, and so on. Moreover, for a $k$-simplex, the boundary operator $\partial_{k}: C_{k} \rightarrow C_{k-1}$ maps each $k$-simplex to the alternating formal sum of its $(k-1)$-dimensional faces, and is defined as follows:

\begin{equation*}
	\partial_{k}\left(\sigma^{k}\right) = \sum_{i=0}^{k} (-1)^{i} \left\{ w_{0}, w_{1}, \cdots, \hat{w}_{i}, \cdots, w_{k} \right\} \tag{1}
\end{equation*}

where $\left\{ w_{0}, w_{1}, \cdots, \hat{w}_{i}, \cdots, w_{k} \right\}$ represents the $(k-1)$-simplex obtained by excluding the $i$-th vertex of the $k$-simplex. This forms a chain complex:

\[\cdots \rightarrow C_{k+1} \xrightarrow{\partial_{k+1}} C_k \xrightarrow{\partial_k} C_{k-1} \rightarrow \cdots.\]

The $k$-cycle group and $k$-boundary group are given by the kernel of $\partial_k$ and the image of $\partial_{k+1}$, respectively: $\operatorname{ker}(\partial_k) = \{ \sigma \in C_k \mid \partial_k \sigma = 0 \}$ and $\operatorname{im}(\partial_{k+1}) = \{ \partial_{k+1} \sigma \mid \sigma \in C_{k+1} \}$. Since $\partial_k \circ \partial_{k+1} = 0$, we have $\operatorname{im}(\partial_{k+1}) \subseteq \operatorname{ker}(\partial_k)$, and thus the $k$-th homology group $H_k$ is defined as the quotient of these two groups:

\[H_{k} = \frac{\operatorname{ker}\left(\partial_{k}\right)}{\operatorname{im}\left(\partial_{k+1}\right)}.\]

The rank of $H_{k}$, known as the $k$-th Betti number $\beta_{k}$, indicates the number of independent $k$-dimensional features, such as connected components, cycles, or voids, that are not boundaries of higher-dimensional elements.

This framework allows for the systematic identification and classification of topological features, quantified by Betti numbers, and is widely used in molecular representations across multiple scales, within a given configuration \cite{verri1993use,zomorodian2004computing,edelsbrunner2002topological}. In \textcolor{blue}{\autoref{fig:simplexes}B}, $\beta_{0}$ represents the number of connected components, $\beta_{1}$ is the number of tunnels, loops, or circles, and $\beta_{2}$ is the number of voids or cavities. At the heart of TDA is persistent homology (PH) which monitors the evolution of these topological features through a process called filtration (see \textcolor{blue}{\autoref{fig:simplexes}D}). Filtration introduces a parametrization to homology, enabling the analysis of topological features as a parameter—typically related to spatial scale or distance—is varied. This process involves creating a sequence of nested subcomplexes $S_{i}$, each representing the state of the molecular system at a different spatial resolution for a simplicial complex $S$ as follows,

$$
\emptyset=S_{0} \subseteq S_{1} \subseteq \cdots \subseteq S_{n}=S.
$$

Moreover, a visual representation of these features through the persistence barcodes \cite{ghrist2008barcodes} provides an intuitive way to further comprehend the structure of the given simplicial complex \cite{chintakunta2015entropy}. The starting point of a bar in a barcode representation indicates the 'birth' of a feature—its first appearance during the filtration process—whereas the endpoint indicates the 'death' of a feature, expressing either its disappearance or merging with another feature. In homological terms, the 'disappearance' of a feature occurs when it becomes the boundary of a higher-dimensional chain. The length of a bar signifies the 'persistence' of the feature. It should be noted here that short-range interactions are characterized by the simplicial complexes generated at small filtration values, while complexes with larger filtration values depict long-range interactions (see \textcolor{blue}{\autoref{fig:simplexes}C}).

Various types of simplicial complexes include Vietoris-Rips (VR) complexes, alpha complexes, \v{C}ech complexes complexes, and clique complexes. In graph theory, a $k$-clique in an undirected graph $G=(W, E)$ is a set of vertices $\sigma=$ $\left\{w_{0}, w_{1}, \ldots, w_{k-1}\right\}$ such that each pair of vertices $\left(w_{i}, w_{j}\right) \in E$ is adjacent and mutually connected by an edge without a concern for direction. Here, $E$ is the set of unordered pairs of vertices representing the edges with a symmetric connection between any two vertices. A clique complex is a simplicial complex where each $k$-clique corresponds to a $(k-1)$-dimensional simplex. For example, a 2-clique (triangle) with 2 vertices $\left\{w_{0}, w_{1}\right\}$ corresponds to a 1-simplex in the simplicial complex. For a directed graph $G=(W, E)$, a $k$-clique is an ordered subset of vertices $\sigma=\left(w_{0}, w_{1}, \ldots, w_{k-1}\right)$ such that for $i<j$, we have $\left(w_{i}, w_{j}\right) \in E$ and $E$ consists for directed edges between the vertices having an asymmetric connection. However, not every set of directed edges between three vertices forms a directed 2-clique. For example, a graph with vertices \( \{w_0, w_1, w_2\} \) and directed edges \( (w_0, w_1), (w_1, w_2), (w_2, w_0) \) does not form a directed 2-clique, because the edge \( (w_0, w_2) \) is missing. For a 2-clique to exist, all necessary directed edges between the vertices must be present. This configuration corresponds to a 3-simplex. The set of all cliques in a directed graph is called a directed clique complex, commonly known as a directed flag complex. 

Within the framework of directed graphs (digraphs), a directed flag complex $d F l(G)$ extends the concept of flag complexes by representing the structure of interactions based on the directionality of edges. In $d F l(G), k$-simplices are entirely comprised of $k+1$ ordered vertices $W$, where the directed edges $E$ exist between each consecutive pair of vertices following the order. Moreover, every directed flag complex is an ordered simplicial complex however, not all ordered simplicial complexes are directed flag complexes because every $k+1$-clique needs to be included in order to form a directed flag complex. This means that an ordered simplicial complexes which are not directed flag complexes can be generated by the excluding cliques with more than two vertices i.e., triangles or higher-order cliques \cite{jones2023persistent}. Persistent Betti numbers provide limited information about the geometric and topological evolution of data. To capture these changes more comprehensively, persistent topological Laplacians offer a deeper analysis. The persistent path Laplacian \cite{wang2023persistent} bridges spectral geometry and topological persistence by tracking the shape and structure of directed graphs through a filtration process. It creates a sequence of Laplacian matrices that reflect changes at various filtration levels. Extending this framework, the topological hyperdigraph Laplacian \cite{chen2023persistent} applies to hypergraphs, which represent complex, asymmetric interactions. This Laplacian captures both harmonic and non-harmonic spectra, paired with persistent homology to monitor the topological evolution of structured data over multiple scales. Lastly, the persistent sheaf Laplacian \cite{wei2021persistent} unifies the spectral theory with cellular sheaves, encoding both geometric and non-geometric characteristics of point clouds. This technique facilitates the amalgamation of diverse data types, delivering richer insights into the foundational structure of the data.

PDFL extends the concept of combinatorial Laplacian to the persistent directed flag complex setting allowing for a more intricate analysis of spectral properties across various filtration levels \cite{jones2023persistent}. While directed flag Laplacian serves as a significant tool for analyzing the spectral properties of directed simplicial complexes, it does not capture topological features at different scales in the context of real-world applications. This is where PDFL plays its part by extending the directed flag Laplacian through the incorporation of the concept of persistence, which tracks the evolution of topological features across various filtration levels.  In particular, PDFL refines the same combinatorial framework of the directed flag Laplacian with an additional layer of persistence, where there is a filtration of directed flag complexes \( \mathrm{dFl}(G_0) \subset \mathrm{dFl}(G_1) \subset \cdots \). This enables the study of topological features across multiple scales. Moreover, the chain groups (or vector spaces) \( C_k \) are defined with coefficients in \( \mathbb{R} \) and are endowed with an inner product. This structure introduces a persistent boundary operator \( \delta_k^{a, b} \) associated with the filtration levels \( a \) and \( b \) for each \( k \geq 0 \) and \( a \leq b \), along with its adjoint \( \left( \delta_k^{a, b} \right)^* \), induced by the inner product structure. These elements combine to form the \( k \)-th \( (a, b) \)-persistent directed flag Laplacian \( \Delta_k^{a, b}: C_k^a \rightarrow C_k^b \), defined as follows:

\begin{equation*}
\Delta_{k}^{a, b}=\delta_{k+1}^{a, b} \circ\left(\delta_{k+1}^{a, b}\right)^{*}+\left(\delta_{k}^{a}\right)^{*} \circ \delta_{k}^{a} \tag{2}
\end{equation*}
with the persistent boundary operator defined as
\begin{equation*}
\delta_{k+1}^{a, b}=\left(i_{k}^{a, b}\right)^{*} \circ \delta_{k+1}^{b} \circ \iota_{k+1}^{a, b} \tag{3}.
\end{equation*}

Here, \( \delta_{k+1}^{a, b} \) is expressed in terms of the inclusion morphisms \( i_{k}^{a, b}: C_{k}^{a} \rightarrow C_{k}^{b} \) and \( \iota_{k+1}^{a,b}: C_{k+1}^{a,b} \rightarrow C_{k+1}^{b} \), where \( C_{k+1}^{a,b} = \{ \sigma \in C_{k+1}^b \mid \delta_{k+1}^b \sigma \in C_k^a\} \). This means that \( C_{k+1}^{a,b} \) consists of chains in \( C_{k+1}^b \) whose boundaries lie entirely within \( C_k^a \). These inclusion morphisms are crucial in the construction of persistent homological structures and allow us to relate chain complexes at various filtration levels. In addition, the adjoints of the boundary operators and inclusion morphisms follow a similar structure. They play a fundamental role in ensuring that the Laplacian is positive semi-definite and self-adjoint by preserving the inner product structure of the chain groups. With $B_{k}^{a}$ and $B_{k+1}^{a, b}$ representing the boundary matrices corresponding to $\delta_{k}^{a}$ and $\delta_{k+1}^{a, b}$, respectively, the matrix representation of the positive semi-definite persistent Laplacian $L_{k}^{a, b}$ is given by

\begin{equation*}
L_{k}^{a, b}=B_{k+1}^{a, b}\left(B_{k+1}^{a, b}\right)^{T}+\left(B_{k}^{a}\right)^{T} B_{k}^{a} \tag{4}.
\end{equation*}

The spectrum $\left(\lambda_{1}\right)_{k}^{a, b},\left(\lambda_{2}\right)_{k}^{a, b}, \ldots,\left(\lambda_{n}\right)_{k}^{a, b}$, arranged in non-decreasing order, consists of real and non-negative eigenvalues. These eigenvalues are crucial for understanding the spectral properties of the complex and for tracing how persistent topological features develop across various filtration scales. In particular, the smallest nonzero eigenvalue of $L_{k}^{a, b}$ is especially significant and is denoted by $\lambda_{k}^{a, b}$. A minimal example of a directed 4-clique, denoted by \( G\), is presented in \textcolor{blue}{\autoref{fig:simplexes}E}. The spectra of the corresponding Laplacian matrices \( L_0 \), \( L_1 \), \( L_2 \), and \( L_3 \) is calculated (see \textcolor{blue}{\autoref{fig:simplexes}F}) which results in Betti numbers \( \beta_0 = 1 \), \( \beta_1 = 0 \), \( \beta_2 = 0 \), and \( \beta_3 = 0 \), respectively. 

 \vspace{1em}    
\subsection{Computational Implementation: Correlation Functions and Protein-Ligand Analysis}

By abstracting intricate topological features and reducing dimensionality, PH unveils the fundamental connectivity within biomolecular complexes and enables accurate predictions of protein-ligand binding affinity. Topological fingerprints (TFs) \cite{xia2014persistent,cang2015topological} play a pivotal role in this framework by providing a detailed characterization of the spatial and interaction patterns within biomolecular complexes. These TFs encompass various topological invariants and are typically visualized using barcodes \cite{ghrist2008barcodes} or persistence diagrams. While TFs can capture connectivity patterns as the filtration parameter varies and identify persistent topological structures, they often overlook critical atomic-level details. To tackle this constraint, we employ Element-Specific Topological Fingerprints (ESTFs) \cite{cang2018integration}, which take into account the distinct elemental composition of proteins and ligands, facilitating a more refined characterization of biomolecular interactions. We represent these interactions by topological connections between atom pairs, denoted as (P - L), where P refers to a protein atom and L refers to a ligand atom. This notation captures the general connectivity between the two entities and provides a biologically relevant analysis of binding interactions. For instance, a carbon-oxygen (\text{C - O}) graph represents interactions by linking the carbon atom from the protein with the oxygen atom from the ligand (see \textcolor{blue}{\autoref{fig:flowchart}B}).

When studying the 3D structure of biomolecules, such as proteins, we consider \( N \) atoms, each represented by their position as \(\{\mathbf{x}_i \mid \mathbf{x}_i \in \mathbb{R}^3, \, i = 1, 2, \dots, N\}\). The Euclidean distance between any \(i\)-th and \(j\)-th atom is denoted by \( ||\mathbf{x}_i - \mathbf{x}_j|| \). For the quantification of the interaction between atoms based on this distance, we introduce the characteristic distance \( \eta_{ij} = \tau(r_i + r_j) \), where \(r_i\) and \(r_j\) represent the van der Waals radii of atoms \(i\) and \(j\), respectively, and \( \tau >0 \) controls the rate of decay of the interaction strength, efficiently tuning the spatial scale at which atomic interactions are significant. Alongside the topological perspective, it is crucial to quantify atomic interaction strengths beyond relying solely on Euclidean distances. A more refined approach uses decaying radial basis functions, often referred to as kernels, to transform these distances into interaction measures.

The flexibility-rigidity index (FRI), introduced in earlier studies \cite{nguyen2017rigidity, opron2014fast, xia2013multiscale}, offers an effective computational framework for predicting B-factors and evaluating atomic-level flexibility based on atomic coordinates. The FRI formulation incorporates a correlation function, denoted as \(\Phi\), which maps Euclidean distances between atoms to values signifying the strength of their pairwise interactions. The rigidity index \( \mu_i \) and the flexibility index \(f_i\) of atom \(i\) are calculated based on a weighted sum of interactions with all other atoms, using the following expressions:

\begin{equation}
	\mu_i = \sum_{j=1}^{N} \omega_j \Phi(||\mathbf{x}_i - \mathbf{x}_j||; \eta_{ij}), \tag{5}
\end{equation}

\begin{equation}
	f_i = \frac{1}{\mu_i}. \tag{6}
\end{equation}

Here, \(\omega_j\) are the element-specific weights, which are set to 1 for all atoms to reduce the number of free parameters. The function \(\Phi\) is a real-valued, monotonically decreasing correlation function that describes how interactions between atoms diminish with increasing distance. It depends on the Euclidean distance \( ||\mathbf{x}_i - \mathbf{x}_j|| \) and the characteristic distance \( \eta_{ij} \). The behavior of \(\Phi\) is governed by the following properties {\cite{opron2014fast}}:

\begin{equation}
	\Phi(||\mathbf{x}_i - \mathbf{x}_j||; \eta_{ij}) = 
	\begin{cases} 
		1, & \text{as} \ ||\mathbf{x}_i - \mathbf{x}_j|| \to 0 \\ 
		0, & \text{as} \ ||\mathbf{x}_i - \mathbf{x}_j|| \to \infty 
	\end{cases} \tag{7}
\end{equation}

Most commonly used FRI correlation functions include generalized exponential functions defined as:

\begin{equation}
	\Phi^{\text{exp}}_{\kappa,\tau}(||\mathbf{x}_i - \mathbf{x}_j||; \eta_{ij}) = e^{-\left(\frac{||\mathbf{x}_i - \mathbf{x}_j||}{\eta_{ij}}\right)^\kappa}, \quad \kappa > 0 
	\tag{8}
\end{equation}

and generalized Lorentz functions defined as:

\begin{equation}
	\Phi^{\text{lor}}_{\nu,\tau}(||\mathbf{x}_i - \mathbf{x}_j||; \eta_{ij}) = \frac{1}{1 + \left(\frac{||\mathbf{x}_i - \mathbf{x}_j||}{\eta_{ij}}\right)^\nu}, \quad \nu > 0 
	\tag{9}
\end{equation}

The exponential kernel decays rapidly, with the parameter \(\kappa\) controlling the rate of this decay. The Lorentz kernel decays more slowly, making it suitable for long-range interactions, with the parameter \(\nu\) adjusting the sharpness of the decay. 

In the present work, we construct an element-level multiscale weighted rigidity index that captures the total interaction strength between protein (\(P\)) and ligand (\(L\)) atom pairs. This interaction strength is quantified by summing up the contributions from all pairs of protein atoms \(p \in P\) and ligand atoms \(l \in L\), as expressed in the following rigidity index equation:

\begin{equation}
	\begin{aligned}
		\label{eq:10}
		RI^{\Omega}_{\beta,\tau, c}(P - L) &= \sum_{p \in P} \mu^{\Omega}_p = \sum_{p \in P} \sum_{l \in L } \Phi^{\Omega}_{\beta,\tau}(||\mathbf{r}_p - \mathbf{r}_l||; \eta_{pl}), \\
		&\forall \ ||\mathbf{r}_p - \mathbf{r}_l|| \leq c
	\end{aligned}
	\tag{10}
\end{equation}

where $\Omega = E$ or $\Omega = L$ refers to a kernel index. The term \(\mu^{\Omega}_p\) represents the rigidity index for a given protein atom \(p\), which is the sum of its interactions with all ligand atoms \(l \in L\). The total rigidity index \(RI_{\beta,\tau}(P - L)\) sums over all protein atoms to provide a global measure of the rigidity for the entire protein-ligand interaction network within a predefined cutoff distance \(c\), thereby reducing computational complexity.

The proposed multiscale system given in equation \eqref{eq:10} represents a bipartite directed graph, as the interactions (edges) are directed exclusively between protein atoms and ligand atoms, with no interactions within the same group. In \eqref{eq:10}, protein atoms \( P \) and ligand atoms \( L \) form two distinct sets, and the edges between them are weighted using a transformed correlation matrix. These edges represent the spatial interaction and connectivity between a protein atom \( p \) and a ligand atom \( l \), defined as \( 1 - \kappa(p, l) \), where \( \kappa(p, l) = \Phi^{\Omega}_{\beta,\tau} \). The resulting transformed values connect nodes representing a protein atom \( p \) to a ligand atom \( l \). This transformation yields values in the range \( (0, 1] \) and ensures that shorter distances between protein-ligand pairs result in higher interaction strengths, while more distant pairs result in weaker interactions. Furthermore, the directionality of the edges in this graph is determined based on the electronegativity values \( \chi_p \) and \( \chi_l \) of the protein and ligand atoms. We set an edge directed from \( p \) to \( l \) for \( \chi_p < \chi_l \), representing a tendency for electron density to flow towards the more electronegative ligand atom. Conversely, if \( \chi_p > \chi_l \), the edge is directed from \( l \) to \( p \). A bonding environment is taken into account when both atoms share the same value of electronegativity. As an example, we consider the case of a weighted edge $(N-N)$ where both protein and ligand nitrogen atoms have same electronegativity value of 3.04. We, then, look into the bond limit of 1.55 \AA\ between the nitrogen protein atom \( S_p \) and the nitrogen ligand atom  \( S_l \) to determine a close proximity of forming a bond. If bonded, we determine the sum of electronegativity values for the atoms bonded to \( S_p \) and \( S_l \). As a result, an edge is directed from the protein N atom to the ligand N atom if \( S_p < S_l \); otherwise, the edge is directed from the ligand N atom to the protein N atom (i.e., for \( S_l < S_p) \). With the correlation matrix functioning as the foundation for interaction weights and electronegativity facilitating the edge directions, this simple construction enables us to record both the directionality and strength of interactions between protein and ligand atoms. Thus, we simplify data processing by incorporating a detailed network of digraphs into our model, while accuracy is preserved by highlighting the fundamental characteristics to represent interactions between atoms.

 \vspace{1.5em}    
\section{Conclusion}

Persistent Homology (PH) and persistent Laplacian (PL) are well-established strategies in topological data analysis (TDA) for analyzing biomolecular systems. However, these methods overlook factors such as polarization, heterogeneity, and directed behavior in chemical, physical, and biological interactions. To address these limitations, we introduce the persistent directed flag Laplacian (PDFL) model as a novel approach for modeling directed interactions and apply it to predict protein-ligand binding affinities. Besides the confirmed predictive accuracy and reliability of our PDFL model, it delivers simplicity by expecting only raw inputs, without the need for complex data optimization or processing.

Levaraging the fast algorithmic power of the PDFL tool, this multiscale framework generates the underlying topological structures of the system in form of spectra within seconds. This asserts its position as a powerful tool and a vital asset in biomolecular modeling and drug discovery. Conclusively, this work marks the first real-world application of PDFL and introduces the first topological modeling of polarized  interactions in molecular science. The success of PDFL demonstrates its high applicability not only to biomolecular modeling, directed evolution, protein engineering, and drug discovery, but also to data science more broadly.

 \vspace{1em}   
\section*{Data and Code availability}
 All codes needed to evaluate the conclusions in this study are available at \href{https://github.com/mzia-s/PDFL-Score.git}{https://github.com/mzia-s/PDFL-Score}, while the source codes for the PDFL tool can be found at \href{https://github.com/bdjones13/flagser-laplacian/}{https://github.com/bdjones13/flagser-laplacian/.} The PDBbind datasets, can be downloaded from \href{https://www.pdbbind.org.cn}{https://www.pdbbind.org.cn}.
   
\vspace{1em}  
\section*{Supporting Information}
\href{./SI.pdf}{Supplementary Information} is available for supplementary tables, figures, and methods. 

\vspace{1em}  
\section*{Acknowledgments}
This work was supported in part by NIH grants  R01GM126189 and R01AI164266, NSF grants DMS-2052983   and IIS-1900473,    MSU Research Foundation,  and Bristol-Myers Squibb 65109.
	
\vspace{1em}  	
\section*{Author Contributions}
\vspace{0.5em}  
M.Z. performed computational work, analyzed results, drafted the manuscript, and wrote the code. B.J. developed the tool and proofread Section 3. H.F. provided his valuable feedback throughout the results stage and proofread Section 2. G.W. designed and supervised the project and revised manuscript. 

\vspace{1em}  
\section*{Conflict of Interest}
\vspace{0.5em}  
The authors declare no conflict of interest.

 \vspace{1em}  
\bibliographystyle{unsrt}
\bibliography{ref} 

\end{document}